% -------------------------------------------------------------------
%   spp.tex :     	Joint paper Hall-Saad 
%   Title:              Eigenvalue bounds for a class of singular
%                       potentials in N dimensions
%   Final version:     	(16 July 98)  Revised 29 Sept 98
% -------------------------------------------------------------------
 \input psfig.sty 
\newif\iftitlepage  \titlepagetrue
\newtoks\titlepagehead  \titlepagehead={\hfil}
\newtoks\titlepagefoot  \titlepagefoot={\hfil}
\newtoks\runningauthor \runningauthor={\hfil}
\newtoks\runningtitle \runningtitle={\hfil}
\newtoks\evenpagehead \newtoks\oddpagehead
\evenpagehead={\hfil\the\runningauthor\hfil}
\oddpagehead={\hfil\the\runnintitle\hfil}
\newtoks\evenpagefoot \evenpagefoot={\hfil\tenrm\folio\hfil}
\newtoks\oddpagefoot  \oddpagefoot={\hfil\tenrm\folio\hfil}
\headline{\iftitlepage\the\titlepagehead\else\ifodd\pageno\the\oddpagehead
\else\the\evenpagehead
\fi\fi}
\footline={\iftitlepage\the\titlepagefoot
\global\titlepagefalse
\else\ifodd\pageno\the\oddpagefoot
\else\the\evenpagefoot\fi\fi}

\newcount\choiceno
\newcount\probno\probno=-1
\def\prob{\futurelet\next\prob@}
\def\prob@{\ifx\next<\expandafter\prob@@\else\expandafter\prob@@@\fi}
\def\prob@@<#1>{\problem<#1>\everypar={\vglue-20pt\endproblem}}
\def\prob@@@{\problem\everypar={vg;ue-20pt\endproblem}}
\newcount\q  \q=0
\newcount\qq  \qq=0
\def\nref {\global\advance\q by1 \item{\bf\the\q.}}
\def\nrf {\global\advance\qq by1 \eqno{(\the\qq)}}
\magnification=\magstep1
\baselineskip 22 true pt    \parskip=0pt plus 5pt
\parindent 0.2in
\hsize 6.15 true in \hoffset .20 true in
\vsize 8.5 true in \voffset .1   true in
\evenpagehead={\vbox{\line{\tenrm\the\runningtitle\hfil\tenbf Page\ \folio}
\medskip\hrule\bigskip}}
\oddpagehead={\vbox{\line{\tenrm\the\runningtitle\hfil\tenbf Page\ \folio}
\medskip\hrule}}
\evenpagefoot={\hfil} \oddpagefoot={\hfil}
\runningauthor={R. L. Hall {\it et al}}
\runningtitle={\it Eigenvalue bounds $\dots$}
\font\chaptitlefont= cmbx10 at 12pt
\font\sectitlefont= cmbx10 at 10pt

\vglue 0.25 true in
\centerline{\chaptitlefont Eigenvalue bounds for a class of}
\medskip
\centerline{\chaptitlefont singular potentials in $N$ dimensions}
\vskip 0.5 true in
\centerline{Richard L. Hall and Nasser Saad
\footnote{$^\dagger$}{Present address: Department of Mathematics, Faculty of Science, 
Notre Dame University, Beirut, Lebanon}}
\vskip .25 true in
{\leftskip=0pt plus 1fil
\rightskip=0pt plus 1fil
\parfillskip=0pt
\baselineskip 18 true pt
\obeylines
Department of Mathematics and Statistics,
Concordia University,
1455 de Maisonneuve Boulevard West,
Montr\'eal, Qu\'ebec,
Canada H3G 1M8.\par}
\bigskip
\bigskip
\centerline{\sectitlefont Abstract}
\bigskip
\vskip .25 true in

\noindent 
The eigenvalue bounds obtained earlier [J. Phys. A: Math. Gen. {\bf 31} (1998) 963] for  smooth transformations of the form $V(x) = g(x^2) + f({1\over x^2})$ are extended to $N$-dimensions. In particular 
a simple formula is derived which bounds the eigenvalues for the spiked harmonic oscillator potential 
$V(x) = x^2 + {\lambda\over x^\alpha},\alpha>0,\lambda>0$, and is valid for all discrete eigenvalues, arbitrary angular momentum $l$ and spatial dimension $N$. 
\vskip 1 true in
\noindent{\bf PACS } 03.65.Ge

\vfill\eject
\noindent Recently, a simple formula that bounds the eigenvalues $E_n$ of Sch\"or- dinger's equation
$$
-\psi^{\prime\prime}+(g(x^2)+f({1\over x^2}))\psi=E_n\psi,\quad 
\psi(0)=0,\nrf
$$
where 
$g$ and $f$ are two smooth transformations of $x^2$ and 
${1\over x^2}$ respectively, was obtained by the present authors. 
They showed that $E_n$ can be approximated by the expression
$$
\eqalign{
E_n\approx&\min_{s,t>0}\left\{ g(s^2)-
s^2g^{\prime}(s^2)+f({1\over t^2})-{1\over t^2}
{f^{\prime}({1\over t^2})}\right.+ \cr
&\left.\qquad\qquad\sqrt{g^{\prime}(s^2)}(4n+2+\sqrt{4f^{\prime}
({1\over t^2})+1})\right\},\quad n=0,1,2,\dots\cr}\nrf
$$
This formula provides a lower bound $ (\approx\ =\ \geq)$ or an upper bound $(\approx\ =\ \leq)$
to the exact eigenvalues $E_n$ of Eq.(1) according as the 
transformation functions $g$ and $f$ are both convex $(g''>0, f''>0)$ or  both concave $(g''<0, f''<0).$ 
This allowed us, for example, to obtain simple expressions
which bound the spectrum of the  spiked harmonic 
oscillator potential $V(x)=\lambda x^2+
{\mu\over {x^\alpha}}$, $\alpha \geq 1$, $n=0,1,2,
\dots,$ namely
$$E_n\approx\epsilon_n({\hat t})=
(1-{\alpha\over 2}){{\mu}\over {{\hat t}^\alpha}}+2\lambda \hat t^2
+2\sqrt{\lambda}(2n+1)\nrf
$$
where ${\hat t}$ is the real positive root of 
$$2\mu\alpha t^{2-\alpha}-4\lambda t^{4}+1=0.$$
Here $\epsilon_n({\hat t})$ is lower bound to $E_n$ when
 $\alpha>2$ and an upper bound when $\alpha<2.$ The purpose of the present paper is to extend these results to the $N$-dimensional case with arbitrary angular momentum number $l$. 

\noindent We notice first that the exact eigenvalues [3]
$$E_{nl}=\sqrt{\lambda}(4n+2+\sqrt{4\mu+(2l+1)^2}),\quad n=0,1,2,\dots$$
of Schr\"odinger's equation with the Gol'dman and Krivchenkov potential $V(x)=\lambda x^2+{\mu\over x^2}$ in 3-dimensions can  be extended [4] to the $N$ dimensional case by replacing $l$ with $l+{N\over 2}-{3\over 2}$. Indeed, these exact solutions could be generated from the well known solutions of harmonic oscillator potential by two simple transformations: first replace the angular momentum  $l$ in the harmonic oscillator energy expression $\sqrt{\lambda}(4n+2l+3),\ n=0,1,2,\dots$ by $-{1\over 2}+\sqrt{\mu+(l+{1\over 2})^2}$; then replace $l$ with   $l+{N\over 2}-{3\over 2}$. Thus the exact eigenvalues of the $N$-dimensional schr\"odinger equation with the Gol'dman and Krivchenkov potential are
$${E}_{nl}^{(N)}=2\sqrt{\lambda}(2n+1+\sqrt{\mu+(l+N/2-1)^2}),\ n,l=0,1,2,\dots\nrf$$ 

The method used to develop the results Eq.(5) and Eq.(6) of Ref.[1] can now be followed, but instead of formula (2) of Ref.[1] we use Eq.(4) above and obtain
$$E_{nl}^{(N)}\approx\min_{s,t>0}\epsilon_{nl}^{(N)}(s,t)$$
where
$$
\eqalign{
\epsilon_{nl}^{(N)}(s,t)&=\left\{g(s^2)-
s^2g^{\prime}(s^2)+f({1\over t^2})-{1\over t^2}
{f^{\prime}({1\over t^2})}\right.\cr
&+\left.2\sqrt{g^{\prime}(s^2)}(2n+1+\sqrt{f^{\prime}
({1\over t^2})+(l+N/2-1)^2})\right\}.\cr}\nrf
$$

\noindent The case where $g(x^2)=\lambda x^\beta$ and 
$f({1\over {x^2}})={ \mu\over {x^\alpha}}$ implies, from (5), that
$$\eqalign{\epsilon_{nl}^{(N)}(s,t)&=\lambda(1-{\beta\over 2})s^\beta+(1-{\alpha\over 2}){{\mu}\over {t^\alpha}}
\cr
&+
\sqrt{{2\lambda
 \beta s^{\beta-2}}}\biggl(2n+1+
\sqrt{{{\mu\alpha}\over {2t^{\alpha-2}}}+(l+N/2-1)^2}\biggr).\cr}\nrf
$$
In particular, for the spiked harmonic oscillator potential with $\beta=2$ it follows from (5) that the eigenvalue approximation is 
given by 
$$
\epsilon_{nl}^{(N)}(t)=(1-{\alpha\over 2}){{\mu}\over {t^\alpha}}+2\lambda t^2+2\sqrt{\lambda}(2n+1)\nrf
$$
where $t$ is real positive root of
$$2\lambda t^4-\mu\alpha t^{2-\alpha}-2(l+N/2-1)^2=0.\nrf$$
 We now prove that for optimal $t$ there is only one positive real root given by (8). If we let $h(t)=2\lambda t^4-\mu \alpha t^{2-\alpha}-2(l+N/2-1)^2$, then for $\alpha<2$: $h(t)\rightarrow -2(l+N/2-1)^2$ as $t\rightarrow 0$ and $h(t)\rightarrow \infty$ as $t\rightarrow \infty$. On the interval $(0,\infty)$ the function $h(t)$ has only one minimum occurring at 
$$
t_{\min}=\bigg({{\lambda\alpha(2-\alpha)}\over 8\mu}\bigg)^{1\over 2+\alpha}
$$
Consequently for $\alpha<2$, Eq.(8) has only one real positive root. For $\alpha>2$,  $h(t)\rightarrow -\infty$ as $t\rightarrow 0^+$ and to $+\infty$ as $t\rightarrow \infty$. On the interval $(0,\infty)$, $h(t)$ is monotone increasing on the open interval $(0,\infty)$ and we conclude that (8) has only one real positive solution for all $\alpha$. The above discussion leads to the following simple expression for the energy bound approximations for the spiked harmonic oscillator potential valid for all dimensions $N \geq 2,$ arbitrary angular momentum $l \geq 0,$ and $n \geq 0$     
$$\left\{
\eqalign{
&\epsilon_{nl}^{(N)}({\hat t})=(1-{\alpha\over 2}){{\mu}\over {{\hat t}^\alpha}}+2\lambda \hat t^2+2\sqrt{\lambda}(2n+1),\cr
&{\hat t}\ {\rm is\ the\ root\ of}\quad 2\lambda t^4-\mu \alpha t^{2-\alpha}-2(l+N/2-1)^2=0.\cr}\right.\nrf
$$
 In Table 1 we exhibit the upper bounds $E_{00}^U$ obtained by use of formula (7) for dimensions $N=2$ to $10$ with $\alpha=1.9,$ $\lambda =1,$ and $\mu=10$, along with some 
accurate values obtained by direct numerical integration 
of Schr\"odinger's equation.  Similar accurate numerical results
could also be obtained by the use of perturbation methods such as the renormalized 
hypervirial perturbation method of Killingbeck~[5].
In Table 2 we exhibit the corresponding lower 
bounds $E_{21}^L$ obtained by use of formula (7) for dimensions $N=2$ to $10$ with $\alpha=2.1,$ $\lambda =1,$ and $\mu=10.$

For the particular test problem discussed here, other approximation methods might also be considered.  For example, if we let $E(\alpha)$ represent an eigenvalue of the operator $H(\alpha) = -\Delta + x^2 + \mu x^{-\alpha}$ with $\mu$ {\it fixed}, then  an immediate first-order approximation `formula' is provided by
$$E(\alpha) \approx E(2) + (\alpha - 2)E'(2).\nrf$$
The problem now is to find $E'(2).$   Since $E(\alpha) = (\psi(\alpha),H(\alpha)\psi(\alpha)),$ differentiation with respect to $\alpha$ and the minimal property of the expectation $(\psi(\alpha),H(2)\psi(\alpha))$ with respect to $\alpha$ leads to the expression:
$$E'(2) = -\mu(\psi(2),\log(x)x^{-2}\psi(2)).\nrf$$
As an illustration of this result we consider the first and last lines of Table~1. We find for $N = 2$ that $E'(2) \approx -1.557$ and $E(1.9) = 8.4803;$ meanwhile for $N = 10,$ $E'(2) \approx -1.498$ and $E(1.9) \approx 12.3479.$ The same reasoning and method can be applied to Table~2.  However, these results are particular to the example chosen for illustration, and they do not in general provide energy {\it bounds.}   Given the correct convexity of the transformation functions, the geometrical methods described in the present paper provide energy bounds on all the discrete eigenvalues in all dimensions $N \geq 2.$

By extending the scope to $N$ dimensions we have generalized our simple general eigenvalue approximation formulae for the 
potential
$$V(x)=g(x^2)+f({1\over x^2}),$$
where $g$ and $f$ are smooth monotone transformations of $x^2$ and $1\over x^2$ respectively.  These results may be used for exploratory purposes and also for seeding direct numerical 
methods. In Figs.(1) and (2) we show the potential, the eigenvalues, and the unnormalized radial wave functions corresponding to the data in Tables (1) and (2).  The computation of such results is greatly helped by {\it a priori} knowledge of the approximate location of the eigenvalues.
\bigskip
\noindent{\sectitlefont Acknowledgment}
\medskip
Partial financial support of this work under Grant No. GP3438 from the Natural 
Sciences and Engineering Research Council of Canada is gratefully 
acknowledged.
\medskip
\vfill\eject
\noindent{\sectitlefont References}

\medskip

{\parindent=20pt
\item{1.}Hall R L and Saad N 1998 J. Phys. A: Math. Gen. {\bf 31} 963.
\item{2.}Gol'dman I I and Krivchenkov V D 1961 {\it Problems in Quantum mechanics}
(London: Pergamon).
\item{3.}Landau L D and Lifshitz E M {\it Quantum Mechanics: Non relativistic theory} (Oxford, Pergamon, 1977). 
\item{4.}H. Mavromatis, {\it Exercises in Quantum Mechanics} (Kluwer Academic press, Dordrecht, 1991).
\item{5.}  Killingbeck J 1981 J. Phys. A: Math. Gen. {\bf 14}
1005.

\par} 
\bigskip
\vfil\eject
\noindent Table(1):  Upper bounds $E_{00}^U$ using (7) for 
$H=-\Delta+x^2+{10\over x^{1.9}}$ for dimension $N=2$ to 10. The `exact' values
$E_{00}$ were obtained by direct numerical integration of Schr\"odinger's equation.

\bigskip 
\bigskip
\hskip 1 true in
\vbox{\tabskip=0pt\offinterlineskip
\def\tablerule{\noalign{\hrule}\noalign{\medskip}}
\halign to200pt{\strut#&#\tabskip=1em plus2em&
\hfil#\hfil&#& \hfil#\hfil&#&
\hfil#\hfil& #\tabskip=0pt\cr\noalign{\medskip}\tablerule&&$N$&&\bf $E_{00}$&&\bf
$E_{00}^U$ &
\cr\noalign{\medskip}\tablerule
\noalign{\medskip}&&2&& $8.485\ 38$&& $8.511\ 90$&  \cr\noalign{\bigskip}
&&3&&  $8.564\ 36$&& $8.590\ 21$& \cr
\noalign{\bigskip}
&&4&&  $8.795\ 44$&& $8.819\ 47$& \cr
\noalign{\bigskip}
&&5&& $9.163\ 09$&& $9.184\ 61$& \cr
\noalign{\bigskip}
&&6&&  $9.646\ 70$&&$9.665\ 48$& \cr
\noalign{\bigskip}
&&7&& $10.225\ 05$&&$10.241\ 20$& \cr
\noalign{\bigskip}
&&8&& $10.879\ 08$&&$10.892\ 89$& \cr
\noalign{\bigskip}
&&9&&$11.592\ 98$&& $11.604\ 78$& \cr
\noalign{\bigskip}
&&10&& $12.354\ 18$&&$12.364\ 29$& \cr
\noalign{\medskip}\tablerule
}}
\vfil\eject
\noindent Table(2):  lower bounds $E_{21}^L$ using (7) for 
$H=-\Delta+x^2+{10\over x^{2.1}}$ for dimension $N=2$ to 10. The `exact' values
$E_{21}$ were obtained by direct numerical integration of Schr\"odinger's equation.

\bigskip 
\bigskip
\hskip 1 true in
\vbox{\tabskip=0pt\offinterlineskip
\def\tablerule{\noalign{\hrule}\noalign{\medskip}}
\halign to200pt{\strut#&#\tabskip=1em plus2em&
\hfil#\hfil&#& \hfil#\hfil&#&
\hfil#\hfil& #\tabskip=0pt\cr\noalign{\medskip}\tablerule&&$N$&&\bf $E_{21}^L$&&\bf
$E_{21}$ &
\cr\noalign{\medskip}\tablerule
\noalign{\medskip}&&2&& $16.457\ 73$&& $16.543\ 63$&  \cr\noalign{\bigskip}
&&3&& $16.826\ 41$&& $16.904\ 44$& \cr
\noalign{\bigskip}
&&4&& $17.312\ 54$&& $17.381\ 71$& \cr
\noalign{\bigskip}
&&5&& $17.895\ 07$&& $17.955\ 44$& \cr
\noalign{\bigskip}
&&6&& $18.554\ 81$&& $18.607\ 07$& \cr
\noalign{\bigskip}
&&7&& $19.275\ 58$&& $19.320\ 69$& \cr
\noalign{\bigskip}
&&8&& $20.044\ 44$&& $20.083\ 41$& \cr
\noalign{\bigskip}
&&9&&$20.851\ 25$ &&$20.885\ 02$& \cr
\noalign{\bigskip}
&&10&& $21.688\ 22$&& $21.717\ 61$ & \cr
\noalign{\medskip}\tablerule
}}

\vfil\eject

\hbox{\vbox{\psfig{figure=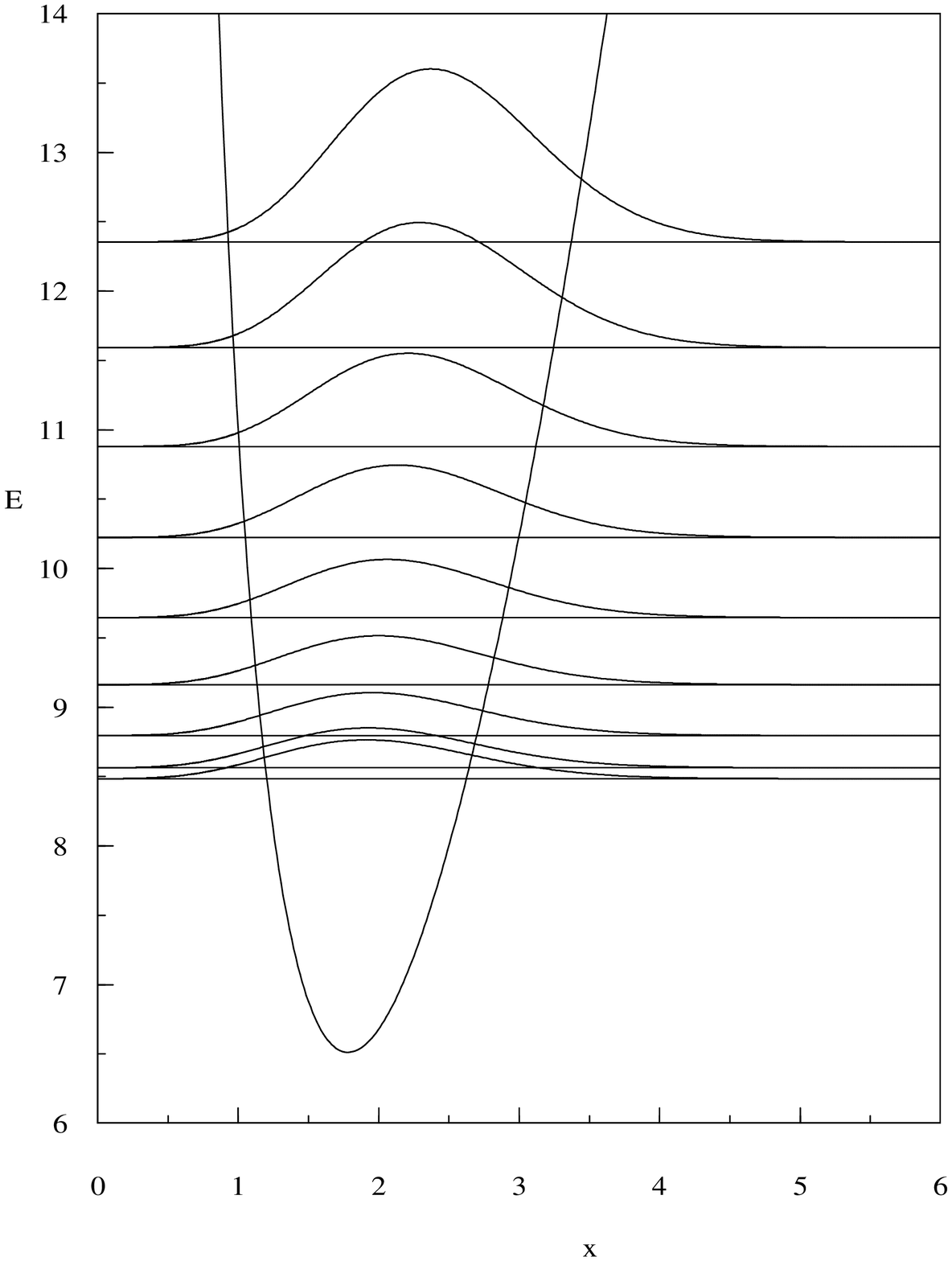,height=6in,width=6in,silent=}}}
\noindent Fig.(1)~~Graph of the eigenvalues $E=E_{00}$ for the Schr\"odinger equation with the potential
 $V(x)=x^2+{10\over x^{1.9}}$ and corresponding unnormalized wavefunctions in dimensions $N=2$ (bottom) to $10$ (top).
\vfil\hfil\break

\hbox{\vbox{\psfig{figure=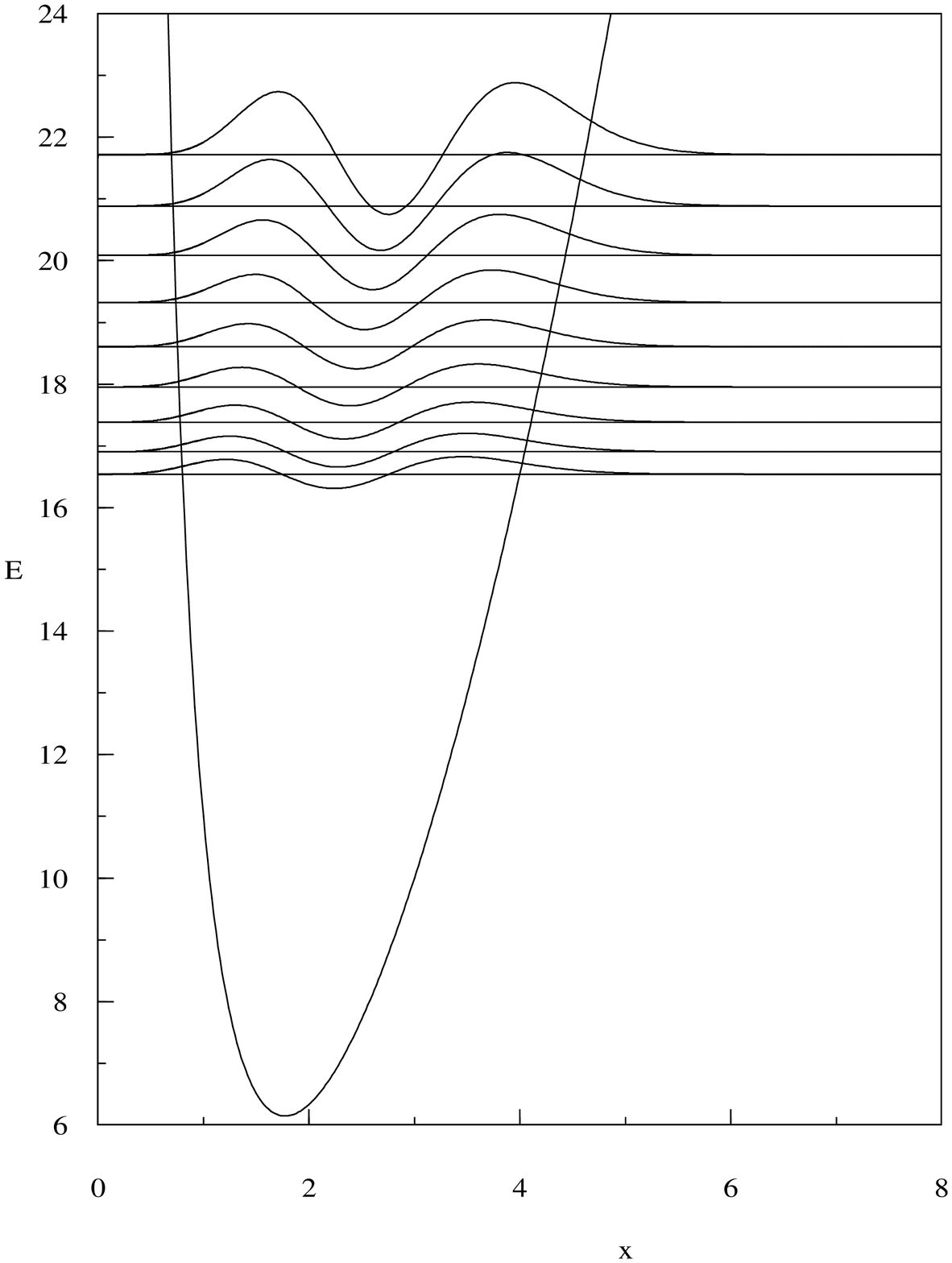,height=6in,width=6in,silent=}}}
\noindent Fig.(2)~~Graph of the eigenvalues $E=E_{21}^L$ for the Schr\"odinger equation with the potential
 $V(x)=x^2+{10\over x^{2.1}}$ and corresponding unnormalized wavefunctions in dimensions $N=2$ (bottom) to $10$ (top).
\end